\documentclass[12pt,preprint]{aastex}
\usepackage{epsfig}

\begin{document}
\title{~~\\ ~~\\ J16021+3326: New Multi-Frequency Observations of a Complex Source}
\shorttitle{J16021+3326}
\author{ S. E. Tremblay\altaffilmark{1},  G. B. Taylor\altaffilmark{1,2},   J. L. Richards\altaffilmark{3}, A. C. S. Readhead\altaffilmark{3}, J. F. Helmboldt\altaffilmark{4}, R. W. Romani\altaffilmark{5}, and S. E. Healey\altaffilmark{5}}

\altaffiltext{1}{Department of Physics and Astronomy, University of New Mexico, Albuquerque, NM 87131; tremblay@unm.edu}
\altaffiltext{2}{National Radio Astronomy Observatory, Socorro NM 87801}
\altaffiltext{3}{Astronomy Department, California Institute of Technology, Pasadena,   CA 91125}
\altaffiltext{4}{Naval Research Laboratory, Code 7213, Washington, DC 20375}
\altaffiltext{5}{Department of Physics, Stanford University, Stanford, CA 94305}

\begin{abstract}
We present multifrequency Very Long Baseline Array (VLBA) observations of J16021+3326. These observations, along with variability data obtained from the Owens Valley Radio Observatory (OVRO) candidate gamma-ray blazar monitoring program, clearly indicate this source is a blazar. The peculiar characteristic of this blazar, which daunted previous classification attempts, is that we appear to be observing down a precessing jet, the mean orientation of which is aligned with us almost exactly.
\end{abstract}



\section{Introduction}


The galaxy J16021+3326 (B1600+335) is classified as a gigahertz peaked spectrum (GPS) radio source (Labiano et al. 2007). GPS sources are so named due to their spectral energy distribution (SED) having a radio turnover frequency around 1 GHz. This class is principally composed of two types of sources: compact symmetric objects (CSOs) and core-jets (e.g., blazars). CSOs are dual-lobed sub-kiloparsec scale sources oriented close to the plane of the sky whose emission is dominated by the hotspots where their jets ram into the surrounding medium. The common opinion is that synchrotron self-absorption within the hotspots causes a spectral turnover around 1 GHz (e.g., de Vries et al. 2009), although free-free absorption has also been invoked (e.g., Bicknell 2003). Core-jets can be classified as GPS sources due to their highly variable SED (Tornikoski et al. 2009), or if by chance the emission is dominated by jet structures of the appropriate size to have a peak around 1 GHz (Scott \& Readhead 1977).  Discriminating the reason a source is a GPS typically requires multi-frequency imaging via Very Long Baseline Interferometery (VLBI) techniques, using instruments such as the Very Long Baseline Array (VLBA), to morphologically classify the source. In principle, long term SED variability studies could also be used to discern why a particular source is classified as a GPS (Tornikoski et al. 2009).

Previous VLBI observations of J16021+3326 by Kulkarni \& Romney (1990) and Dallacasa et al. (1995) were unable to disentangle the source structure well enough to explain why this is a GPS source. After observations at 5 GHz in the VLBA Imaging and Polarimetry Survey (VIPS, Helmboldt et al. 2007), we decided to perform multi-wavelength followup with the VLBA to clarify the nature of this source.

In this paper, we have adopted a redshift for J16021+3326 of 1.1\footnote{Since there are no spectroscopic data cited within these papers nor within their references, this redshift should only be taken as a possible value and spectroscopic observations should be published with a verifiable value of z} as per Snellen et al. (2000) \& Labiano et al. (2007). Throughout this discussion, we assume H$_{0}$=73 km s$^{-1}$
Mpc$^{-1}$, $\Omega_m$ = 0.27, $\Omega_\Lambda$ = 0.73,  so 1 mas = 7.991 pc.


\section{Observations and Data Reduction}
\label{observations}

Multi-frequency observations of J16021+3326 were performed on February 19, 2007 with the VLBA,  a summary of these observations is presented in Table \ref{Observations}. These observations consisted of four 8 MHz wide IFs in the C, X and U bands with full polarization centered at: 4605.5 , 4675.5, 4990.5, 5091.5, 8106.0, 8176.0, 8491.0, 8590.0, 14902.5, 14910.5, 15356.5 and 15364.5 MHz at an aggregate bit rate of 256 Mbps to maximize ($u$,$v$) coverage and sensitivity. When the data in each band were combined, the three central frequencies were: 4844.7, 8344.7, and 15137.5 MHz. The integrations were performed in blocks ($\sim$2 minutes for 5 and 8 GHz, $\sim$7.5 minutes for 15 GHz) and these blocks were spread out over a 10 hour period to maximize ($u$,$v$) coverage of the source.

Most of the calibration and initial imaging of the data were carried out by automated AIPS \citep{2003ASSL..285..109G} and Difmap \citep{1997ASPC..125...77S} scripts similar to those used in reducing the VIPS 5 GHz survey data \citep{2007ApJ...658..203H, 2005ApJS..159...27T}. To summarize, flagging of bad data and calibration were performed using the VLBA data calibration pipeline \citep{2005ASPC..340..613S}, while imaging was performed using Difmap scripts described in \citet{2005ApJS..159...27T}. Final imaging was performed manually using the Difmap program, with beam sizes of $2.016\times2.597$ mas in position angle -31.86$^\circ$, $1.176\times1.557$ mas in position angle -32.85$^\circ$ and $0.6284\times0.979$ mas in position angle -29.68$^\circ$ for 5, 8 and 15 GHz respectively. For polarimetry purposes, the lower two and upper two IFs in each band were paired and imaged.

J16021+3326 has also been observed regularly at 15~GHz with the 40~m telescope at the Owens Valley Radio Observatory (OVRO) since mid-2007 as part of an ongoing candidate gamma-ray blazar monitoring program. This program monitors all 1158 sources with declination greater than 20$^\circ$ in the CGRaBS sample (Healey et al., 2008) at least twice per week when weather conditions permit. The CGRaBS sample consists of radio sources selected as likely to be active gamma-ray emitters based on their flat radio spectra and X-ray fluxes.  CGRaBS sources are predominantly known blazars, but also include a few other types. J16021+3326 is identified as a narrow-line radio galaxy in the CGRaBS catalog. 

The OVRO 15~GHz flux densities of J16021+3326 were measured using azimuth double switching as described in Readhead et al. (1989).  The relative uncertainties in flux density result from a 5~mJy typical thermal uncertainty in quadrature with a 1.4\% to 2\% non-thermal random error contribution.  The absolute flux density scale is calibrated to about 5\% using the Baars et al. model for 3C~286 (Baars et al., 1977).  This absolute uncertainty is not included in the plotted errors. In addition to several interruptions in the monitoring program due to hardware failures, J16021+3326 was one of a subset of the sample affected by an unreliable pointing procedure during the period MJD~54750--54908 which resulted in spurious fluctuations in the measured flux densities.  All the potentially significant fluctuations in this period were closely correlated with similar fluctuations in other unrelated sources, suggesting the intrinsic flux density of J16021+3326 was relatively constant during this interval.

An optical r-band observation of J16021+3326 was also performed, using five 300 second exposures with the Palomar 200" telescope in July 2008.


\section{Results}
\label{results}

\subsection{Images}
\label{images}
\begin{figure}
\epsscale{0.4}
\plotone{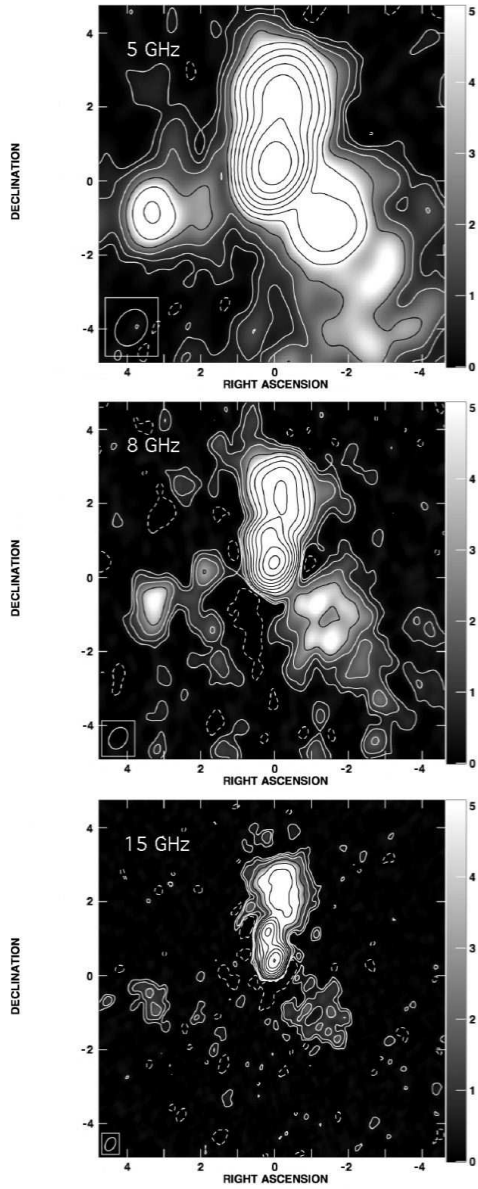}
\caption{  A zoomed in view of the VLBA observations from 2007 February of J16021+3326 at frequencies of (\textit{from top to bottom}) 4.845, 8.345, and 15.137 GHz. Contour levels begin at 0.35 mJy, 0.4 mJy and 0.3 mJy respectively and increase by factors of $2^{1/2}$. The grayscale shows the details of the emission between 0 and 5 mJy. 
}
\label{zoom}
\end{figure}

Figures \ref{images} and \ref{zoom} show the 5, 8, and 15 GHz VLBA images made from the 2007 February observations of J16021+3326. In Figure \ref{images} all three images have been convolved with identical beams, both to enable simpler comparison between the maps and to highlight the detected diffuse emission at 15 GHz. The scale of Fig. \ref{zoom} was chosen to focus on the structure surrounding the phase-center. All three images in Fig. \ref{zoom} are displayed at their natural resolutions to better show the complicated small-scale structure. Self-calibration was utilized when reducing these data, so all absolute positions are lost since the brightest emission was placed at the phase center. The 5 GHz images reveal a northern elongation (B) coming from the phase center (A), and more diminished spurs towards the east (D) and southwest (E). Diffuse emission is seen $\sim$ 40 mas to the east (F), as well as surrounding the southwest spur (E). In the 8 GHz image the diffuse emission is much less prominent  but components A, B, C \& D are all still easily identifiable, although the southwest spur (C) has resolved into a ring. At 15 GHz, the brightest component (A) has resolved into two distinct components (A1, A2) and the diffuse eastern emission (F) has almost completely disappeared. The eastern spur (D) is barely detected, and the southwestern spur (C) has marginally retained its ring-like structure.

\begin{figure}
\epsscale{0.7}
\plotone{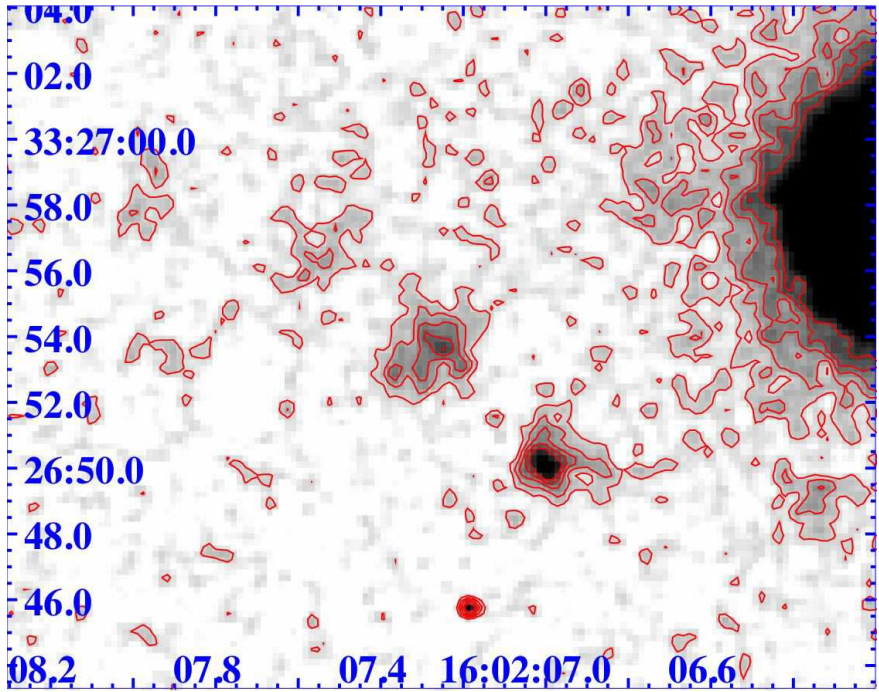}
\caption{Optical r-band image of J16021+3326 (center of field).  This image was taken with the Palomar 200" telescope over five 300s exposures in July 2008. The central 1.1" has a flux density of 0.73 $\mu$Jy, while the entire galaxy, integrated over a 3.6" aperture, has a flux density of 1.4 $\mu$Jy.
}
\label{optical}
\end{figure}

The r-band images obtained by the Palomar 200" (Fig. \ref{optical}) show a faint (1.4 $\mu$Jy integrated over a 3.6"  aperture, for an absolute magnitude of 21.0) galaxy with evidence of a disturbed morphology, which is consistent with previous optical observations by Stickel \& Kuhr (1996). 

\subsection{Spectral Index Distribution}
\label{spectral_index}

\begin{figure}
\epsscale{0.7}
\plotone{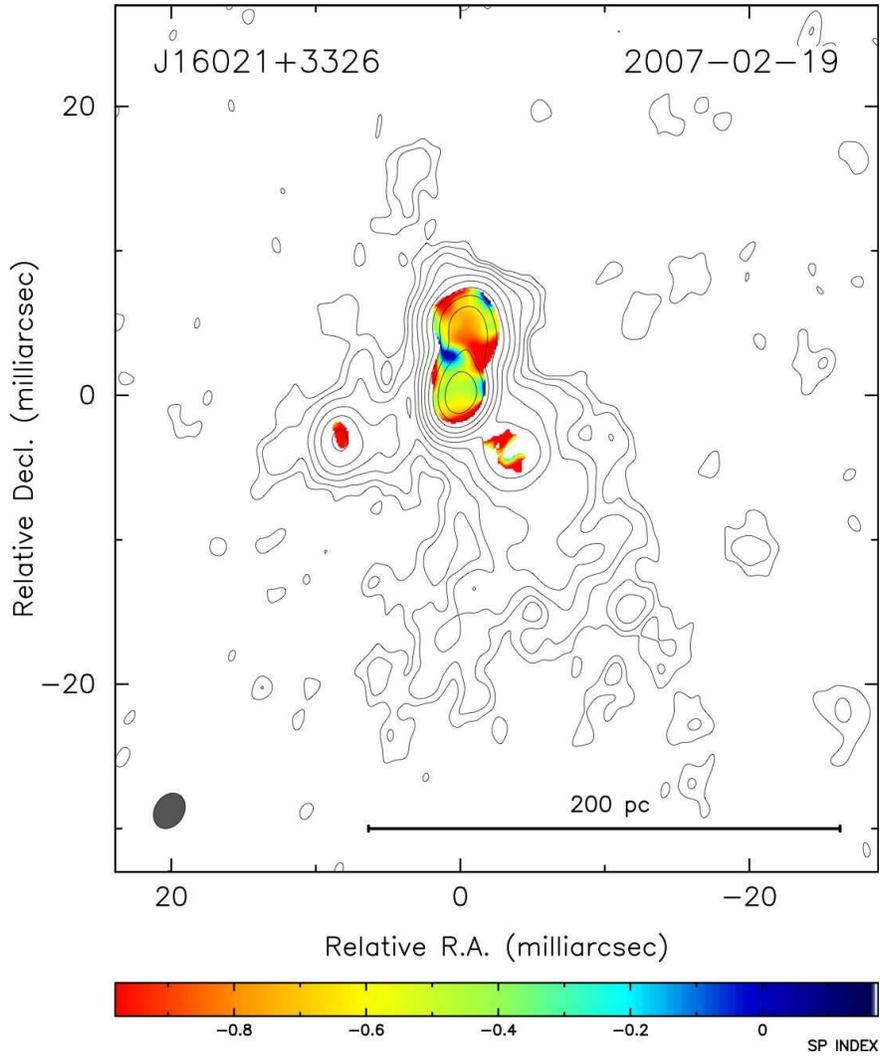}
\caption{Multifrequency observations of J16021+3326; 5 GHz contours overlaid on an 8-15 GHz spectral index image. The flat-spectrum ($\alpha\approx-0.05$) compact component likely identifies the core of the source.  
}
\label{spix}
\end{figure}

The 8 and 15 GHz images were matched in resolution in order to obtain a spectral index distribution across the source, where we take $F_\nu\propto \nu^\alpha$, which was overlaid onto 5 GHz contours to highlight the placement of the distribution within the source structure (Fig. \ref{spix}). Averaged spectral indices for components A1, A2, B, C, D and F and can be found in Table \ref{SpectralIndex}. In summary, the only component exhibiting a flat-spectrum (i.e., an $\alpha \approx 0$) is A2.

\subsection{Polarization}
\label{polarization}

\begin{figure}
\epsscale{0.5}
\plotone{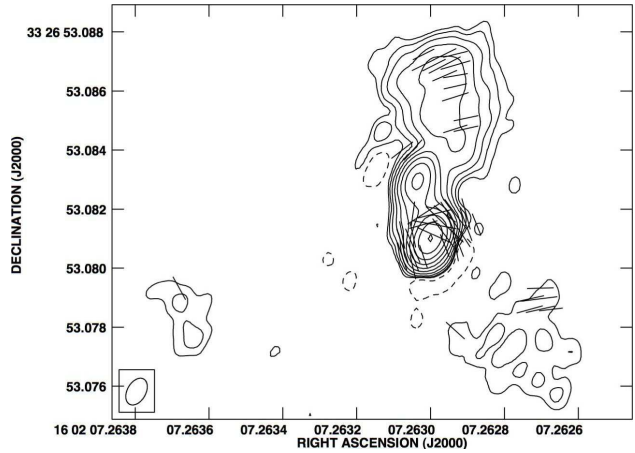}
\caption{Uncorrected polarization electric field (\textit{E}) vectors from one of the two averaged IF pairs at 15 GHz overlaid onto 15 GHz total intensity  contours. The vector lengths are proportional to the polarized flux density, and 1 mas = $7.58\times 10^{-4}$ Jy/Beam. Contour levels begin at 0.6 mJy and increase by factors of $2^{1/2}$. This is an example of how polarization was detected throughout the source, although the exact positions were frequency dependent.
}
\label{upol}
\end{figure}

\begin{figure}
\epsscale{0.6}
\plotone{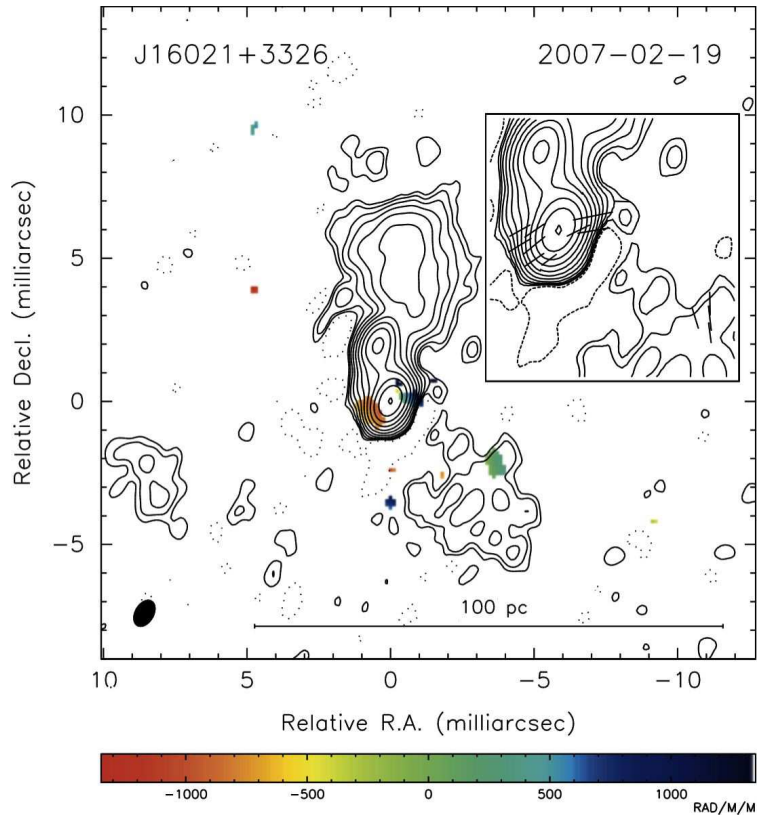}
\caption{Rotation measure map computed from polarization angles measured at three frequencies overlaid onto 15 GHz contours.  Contour levels begin at 0.3 mJy and increase by factors of $2^{1/2}$. The inset shows the corrected polarization angle of the magnetic fields (\textit{B}) of the source.
}
\label{rm}
\end{figure}

Polarization was detected at both 8 and 15 GHz with up to $14.9\pm2.0\%$ and $5.7\pm0.7\%$ polarization respectively. Figure \ref{upol}, showing the polarized flux from the IF pair centered at 15.3605 GHz, provides an example of the polarization maps obtained. Components A, B, C and D all exhibit polarization in the four IF pairs, although the pair centered at 14.9065 GHz detected significantly less polarization the reason for which is unknown. The three IF pairs that detected stronger polarization were used to compute rotation measures ($RM$s) across the source (Fig. \ref{rm}) wherever polarization was detected at all three pairs by fitting the change in polarization angle ($\beta$) to $RM=\frac{\beta}{\lambda^2}$, with values ranging from -984 to 1426 rad m$^{-2}$. The $RM$ is dependent on the magnetic field strength ($B$) and the electron number density ($n_e$) by:
\begin{equation}
 RM=\frac{e^3}{2\pi m_e^2c^4}\int_0^sn_eBds
\end{equation}
where $e$ and $m_e$ are the charge and mass of the electron and $c$ is the speed of light. These rotation measures were then used to calculate the magnetic field polarization angle corrected for Faraday rotation (Fig \ref{rm} inset). It is worth noting that the $RM$ changes sign across A1, which is indicative of a reversal of the magnetic field as seen in the above equation.

\subsection{Variability}
\label{variability}
\begin{figure}
\epsscale{1.0}
\plotone{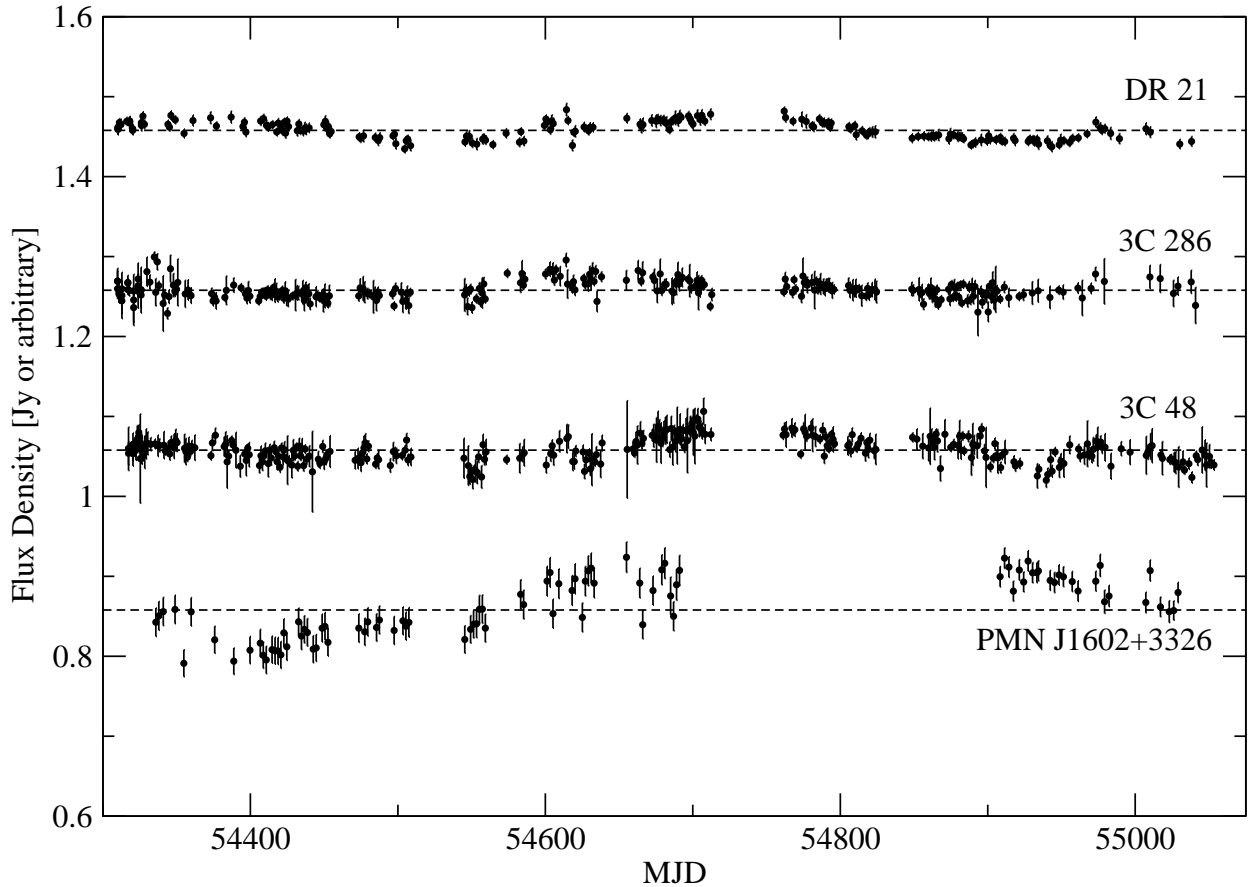}
\caption{
OVRO~40m 15~GHz light curve for J16021+3326 (bottom, in Jy).  Shown for reference are scaled and arbitrarily offset light curves for three stable-flux density sources DR~21, 3C~286, and 3C~48. The reference curves were scaled to the same median flux density as J16021+3326, then offset for display. Dashed lines indicate the median of each light curve. Slow variation seen in the flux density of J16021+3326 (15.5\% peak-to-peak, 4.2\% RMS ) is greater in magnitude than the systematic fluctuations observed in the stable sources (6\%-9\% peak-to-peak, 1.2-1.7\% RMS).
 }
\label{var}
\end{figure}

The OVRO 15~GHz light curve for J16021+3326 is shown in Figure~\ref{var}, along with scaled and offset light curves for the bright, constant-flux sources 3C~48, 3C~286, and DR~21 for reference.  The median flux density of J16021+3326 is 858~mJy and the root mean square (RMS) is 36~mJy, or 4.2\% of the median.  The RMS fluctuations of 3C~48, 3C~286, and DR~21 are 1.7\%, 1.3\%, and 1.2\% of the median flux density, respectively.  The peak-to-peak variation is 133~mJy, or 15.5\% of the median.We conclude that the variation in flux density of J16021+3326 results from actual source variations rather than systematic instrumental fluctuations. These observed fluctuations are consistent with the variability of the flux density (up to 60\% peak-to-peak at 5 GHz over a 6 year period) noted by Dallacasa et al. (1995).


\section{Discussion}

\subsection{Core Identification}
The unresolved, flat-spectrum ($\alpha\sim -0.05$), unpolarized component seen in Fig. \ref{spix} corresponds with A2 (Fig. \ref{zoom}), and is consistent with the emission from the base of the radio-jet, commonly referred to as the core (Begelman et al. 1984). The remainder of the emission surrounding the core (A1, B, C, D, E, and F)  all exhibit steeper spectra ($\alpha\sim -0.54$ to $-1.05$) and are consistent with the identification of these components as jets or hot spots.

\subsection{Morphology}
 Taken alone, the components A1, A2, and B look very much like a CSO, in which the core feeds into two symmetric lobes to the north and south. However, this scenario does not readily explain the remaining components, with the positions of D and F being particularly difficult to reconcile. This source also doesn't resemble a simple core-jet since there are extended components all around the core. Clearly, imaging alone is not enough to determine the proper classification of this complicated source. 
 
 The OVRO variability data argue against J16021+3326 being a CSO. Since the jets associated with CSOs are oriented close to the plane of the sky there is little to no relativistic beaming of their emission expected and the emission is dominated by the extended hotspots where the jets terminate. The flux densities of 5 CSOs were observed to fluctuate over an 8 month period with an average RMS of 0.7\% (Fassnacht \& Taylor 2001), while the two core-jet sources observed over the same period had an average RMS of 3\%. Therefore, the observed variability of this source, 15.5\% peak-to-peak and 4.2\% RMS over a 2 year period, strongly supports the blazar scenario.  

 Lastly, the polarization data also argue against J16021+3326 being a CSO. Since the jets of CSOs are oriented close to the plane of the sky, and are on the size scale of 10s of parsecs they should be viewed through the torus according to AGN unification models (Antonucci 1993). This torus can act as a Faraday screen, thus depolarizing the synchrotron radiation as it leaves the galaxy (Peck \& Taylor 2000; Gugliucci et al. 2007). Very few CSOs have been observed to be polarized, and those that are only exhibit polarization in the oncoming jet. Since we observe polarized flux all around the core of this source, it is unlikely to be a CSO. The derived RMs also lie within the typical blazar range of 500 to several thousand rad m$^{-2}$ (Zavala \& Taylor 2004).

While there seems to be plenty of evidence of the blazar nature of J16021+3326, it is far from a conventional member of the class. Typically, one would expect to see a one sided jet coming out of the core, while in this case there appears to be jet emission all around the core. Precession, observed in other radio sources (e.g., SS 433; Dubner et al. 1998), provides a simple explanation for this unconventional morphology from a blazar. If the jet is directed exceptionally close to the line of sight, any precession would make the jet appear to spiral around the core instead of producing the more familiar core-jet morphology typically exhibited by blazars. The structure observed in J16021+3326 (Fig. \ref{images}) is consistent with knots moving along a spiraling jet and/or foreground hotspot projection.

 J16021+3326 does display two non-blazar-like characteristics since it both appears to lack broad-line emission and it is optically extended (i.e., not a core dominated optical quasar). The first of these appears to be a non-issue since the spectral classification is not well supported in the literature, and yields a second reason a spectrum of this source should be observed and published in the future. As far as the optical classification, there are other examples of sources optically classified as galaxies exhibiting blazar like behavior in the radio (e.g., 3C 111 which exhibits superluminal motion and high variability; Cohen et al. 2007).  It is possible this source is obscured by dust in the optical, but unaffected at radio wavelengths.


\section {Conclusions}

Analysis of multi-frequency (5, 8, and 15 GHz) VLBA data show a compact, flat-spectrum ($\alpha\sim$ -0.05) component that we identify as the core of J16021+3326 surrounded by steeper-spectrum ($\alpha\sim$  -0.5 to -1.0) polarized regions of emission. Also, data from the OVRO candidate gamma-ray blazar monitoring program reveal a relatively high level of variability (15.5\% peak-to-peak and 4.2\% RMS over a two year period). This evidence, coupled with the detection of polarization in most components other than the core, indicate that J16021+3326 is a blazar. The peculiar structure observed is possibly due to the jet, presumably pointed close to the line of sight, precessing around and tracing out a spiral on the sky.

\begin{acknowledgements}
The National Radio Astronomy Observatory is a facility of the National Science Foundation operated under cooperative agreement by Associated Universities, Inc.
\end{acknowledgements}

{\it Facilities:} \facility{VLBA ()}

\bibliographystyle{apj}



\begin{figure}
\epsscale{0.4}
\plotone{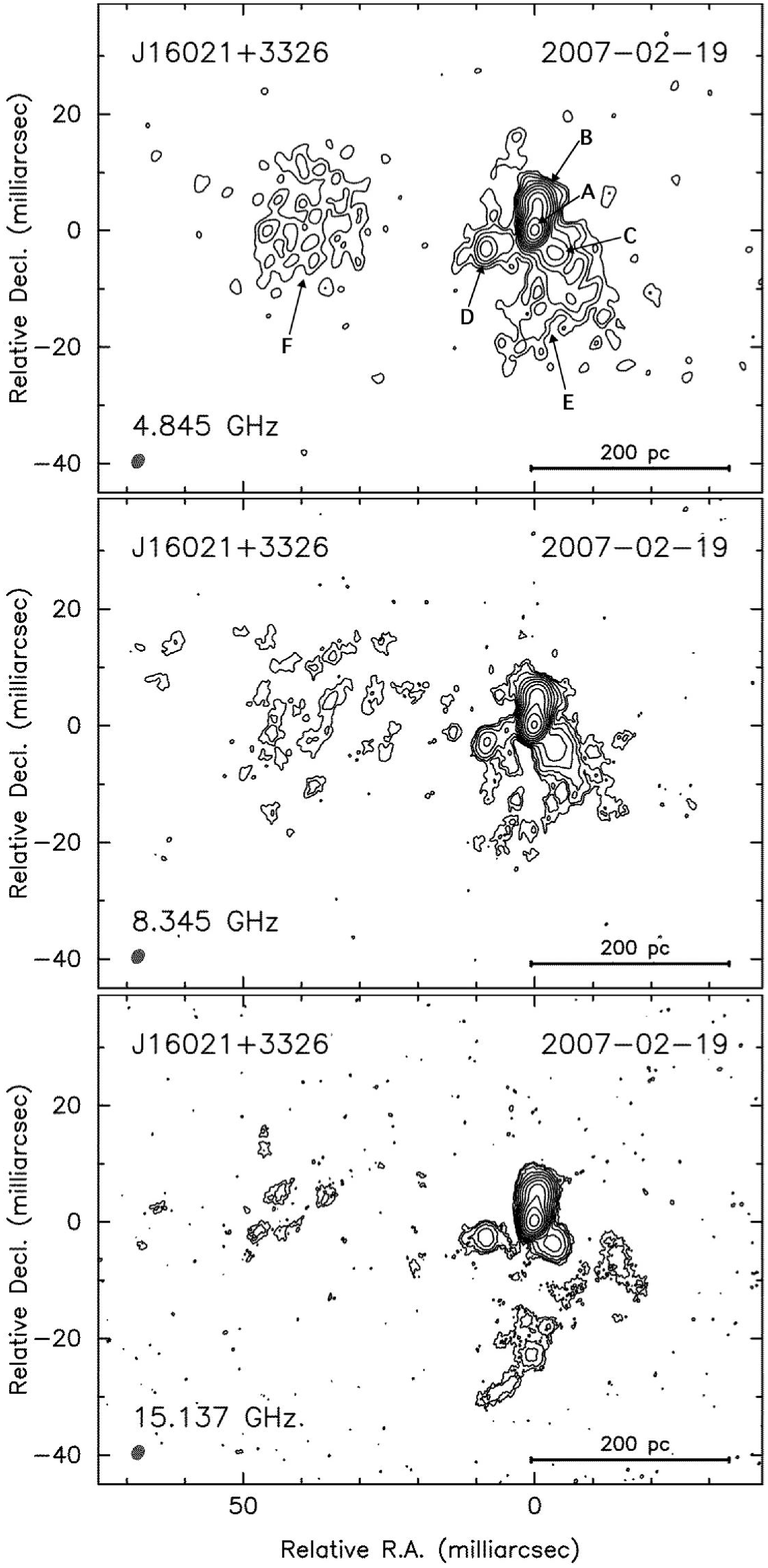}
\caption{VLBA observations from 2007 February of J16021+3326 at frequencies of (\textit{from top to bottom}) 4.845, 8.345, and 15.137 GHz. Contour levels begin at 0.5 mJy, 0.5 mJy and 0.3 mJy respectively and increase by factors of $2^{1/2}$. All three of these images have been convolved with the 5 GHz beam to enable easy comparison at matched resolution. The various components, A through F, have been labeled on the 4.845 GHz image, ordered such that subsequent letters correspond to decreasing peak flux density.
}
\label{images}
\end{figure}

\begin{deluxetable}{lccccccc}
\tabletypesize{\scriptsize}
\tablecolumns{8}
\tablewidth{0pt}
\tablecaption{VLBA Observations of J16021+3326.\label{Observations}}
\tablehead{\colhead{Freq.}&\colhead{Date}&\colhead{Time}&\colhead{BW}
&\colhead{Pol.}&\colhead{IFs}&\colhead{Peak}&\colhead{rms}
 \\
\colhead{(GHz)} & \colhead{} & \colhead{(minutes)} & \colhead{(MHz)} 
& \colhead{}&\colhead{}&\colhead{(mJy beam$^{-1}$)}&\colhead{(mJy beam$^{-1}$)}}

\startdata
4.8447 & 2007 Feb 19 & 24.9 & 32 &4 & 4 &685.06 & 0.15 \\
8.3447 & 2007 Feb 19 &  26.9 & 32 & 4 & 4 & 507.36 & 0.15 \\
15.138& 2007 Feb 19 & 90.8 & 32 & 4 & 4 & 319.62 &  0.11 \\

\enddata

\end{deluxetable}

\begin{deluxetable}{ccc}
\tabletypesize{\scriptsize}
\tablecolumns{3}
\tablewidth{0pt}
\tablecaption{8-15 GHz Spectral Index of Various Components of J16021+3326\label{SpectralIndex}}
\tablehead{\colhead{Component}&\colhead{Spectral Index ($\alpha$)\tablenotemark{\dagger}}&\colhead{RMS\tablenotemark{\ddagger}}}

\startdata
A1 & -0.54 & 0.08\\
A2 & -0.05 & 0.07\\
B & -0.75 & 0.07\\
C & -1.05 & 0.11\\
D & -0.97 & 0.35\\
F & -0.79 & 0.72\\
\enddata
\tablenotetext{\dagger}{Here we use the convention of $F_\nu\propto \nu^\alpha$ and $\alpha$ is averaged over the component using the 8 \& 15 GHz data.}
\tablenotetext{\ddagger}{This is the RMS of $\alpha$ for the averaged area.}

\end{deluxetable}

\end{document}